\documentclass[12pt]{article}

\usepackage{amssymb}
\usepackage{amsmath}
\usepackage{amscd}
\usepackage{latexsym}
\usepackage{graphicx}
\usepackage{url}

\usepackage{cite}

\newcommand{\be}{\begin{equation}}
\newcommand{\ee}{\end{equation}}

\begin{document}

\begin{center}

{\Large{\bf Cancer risk is not (just) bad luck}
 \\ [5mm]

D. Sornette and M. Favre} \\ [3mm]

{\it
Department of Management, Technology and Economics, \\
ETH Z\"urich (Swiss Federal Institute of Technology) \\
Scheuchzerstrasse 7,  Z\"urich CH-8032, Switzerland 
}

\end{center}

\begin{abstract}
Tomasetti and Vogelstein recently proposed that the majority of variation
in cancer risk among tissues is due to ``bad luck," that is, random mutations
arising during DNA replication in normal noncancerous stem cells. They generalize this finding to cancer overall, claiming that ``the stochastic effects of DNA replication appear to be the major contributor to cancer in humans."
We show that this conclusion results from a logical fallacy based on ignoring 
the influence of population heterogeneity in correlations exhibited at the level of the whole population.
Because environmental and genetic factors cannot explain the huge differences in cancer rates between different organs,
it is wrong to conclude that these factors play a minor role in cancer rates. In contrast, we show
that one can indeed measure huge differences in cancer rates between different organs and, at the same time, observe a strong effect of 
environmental and genetic factors in cancer rates. 
\end{abstract}

\vskip 1cm

Tomasetti and Vogelstein showed that the lifetime
risk of cancers of many different types is strongly correlated (0.81) with the
total number of divisions of the normal self-renewing cells maintaining organ-specific
tissue's homeostasis \cite{Tomavogel15}. They conclude from this that the majority of variation
in cancer risk among tissues is due to ``bad luck," that is, random mutations
arising during DNA replication in normal noncancerous stem cells. Generalizing to cancer causation, they claim that ``these stochastic influences are in fact the major contributors to cancer overall, often more important than either hereditary or external environmental factors." In a review by Couzin-Frankel \cite{Couzin-Frankel15} of Tomasetti and Vogelstein's article supported by an interview of Tomasetti, the above mentioned correlation is interpreted as excluding in large part the role of hereditary or environmental factors in the generation of cancers. Couzin-Frankel claims that Tomasetti and Vogelstein's results ``explained two-thirds of all cancers."

Here, we show that this conclusion is fundamentally flawed, as it rests on 
neglecting the influence of population heterogeneity in correlations exhibited at the level of the whole population. Tomasetti and Vogelstein's results 
quantify nicely that a large part of the differences in organ-specific cancer risk can be explained by 
the number of stem cell divisions in different tissues. But the logical fallacy is to extrapolate that, 
because environmental and genetic factors cannot explain the huge differences in cancer rates between different organs,
then these factors play a minor role in cancer rates. In contrast, we show
that one can indeed measure huge differences in cancer rates between different organs and at the same time observe a strong effect of 
environmental and genetic factors in cancer rates.
 
To make our demonstration as clear as possible, we imagine an hypothetical population partitioned
into two groups. The first group exhibits a much lower cancer rate than the second group. 
This may be due to hereditary and environmental factors playing an important role, in addition to 
the number of stem cell divisions in organs. 
We show that, for any given organ, a correlation between lifetime cancer risk and the total number of stem cell divisions at the group level (averaged over the whole population) translates into an equal or higher correlation at the level of the whole population. This, however, says nothing about a possible heterogeneity in susceptibilities to external factors such as genetics or environment. 

For each of the two groups, we assume that the linear correlation of the type
found in Ref. \cite{Tomavogel15} holds:
\be
C_i^{(1)} = \beta^{(1)}  S_i^{(1)} + \epsilon_i^{(1)}~,
\label{rtwhedyhbgw}
\ee
\be
C_i^{(2)} = \beta^{(2)}  S_i^{(2)} + \epsilon_i^{(2)}~.
\label{rtwhedyhbgw2}
\ee
$C_i^{(1)}$ and $C_i^{(2)}$ are the logarithms in base $10$ of the lifetime cancer risks for group 1 and group 2, respectively,
for organ tissue $i$. $S_i^{(1)}$ and $S_i^{(2)}$ are the  logarithms in base $10$ of the total numbers of divisions of stem cells in
group 1 and group 2, respectively, for organ tissue $i$.
$\epsilon_i^{(1)}$ and $\epsilon_i^{(1)}$ are the logarithms in base $10$ of the contributions to lifetime cancer risks
in the two groups in organ tissue $i$ not explained by stem cell divisions.\footnote{Given the range of  lifetime cancer risks 
from $10^{-5}$ to $0.3$ and of the total numbers of divisions of stem cells 
from $10^6$ to $10^{13}$, for a linear correlation analysis (Pearson correlation coefficient), 
Tomasetti and Vogelstein \cite{Tomavogel15} used 
these logarithmic variables (see their supplementary materials). The relevance of the use of 
log-variables is further suggested by their definition of the 
``extra risk score'' \cite{Tomavogel15}.} Finally, the coefficients $\beta^{(1)}$ and $\beta^{(1)}$ quantify the correlation
between $C_i^{(j)}$ and $S_i^{(j)}$, $j=1,2$, across all organ tissues.

The correlation between $C_i^{(j)}$ and $S_i^{(j)}$ is given by
\be
{\rm Corr}[C_i^{(j)}, S_i^{(j)}] := { \beta^{(j)}  {\rm Var}[S_i^{(j)}]  \over
\sqrt{ {\rm Var}[C_i^{(j)}] {\rm Var}[S_i^{(j)}]} }
\label{gbfwb}
\ee
We also introduce the covariance between $C_i^{(j)}$ and $S_i^{(j)}$ defined by
\be 
{\rm Cov}[C_i^{(j)}, S_i^{(j)}] :=  \beta^{(j)}  {\rm Var}[S_i^{(j)}]  ~.
\label{rgheythbwgr}
\ee

The variances of $C_i^{(j)}$ are 
\be
{\rm Var}[C_i^{(j)}] := [\beta^{(j)}]^2  {\rm Var}[S_i^{(j)}] + {\rm Var}[\epsilon_i^{(j)}]~.
\label{vfewfwc}
\ee

We assume that the correlations 
\be
{\rm Corr}[C_i^{(1)}, S_i^{(1)}] = {\rm Corr}[C_i^{(2)}, S_i^{(2)}] := \rho~,
\label{rynehtbgw}
\ee
are the same in both groups, while the incidence of cancers is much higher in the second group.
How is this possible?  To make the example simple, we assume that the rate of
divisions of the normal self-renewing cells maintaining the homeostasis of a given tissue $i$
is approximately the same for all members of our population, and thus the same in both groups. This amounts to assuming 
\be
S_i^{(1)}=S_i^{(2)} := S_i~.
\label{fgngbw}
\ee

To keep our derivation simple, we assume that the logarithm in base 10 of the contribution to lifetime cancer risks not explained by stem cell divisions, namely $\epsilon_i^{(j)}$ ($j=1,2$), has a mean value equal to zero and is solely characterised
by its variance ${\rm Var}[\epsilon_i^{(j)}]$. Then, by definition, the corresponding lifetime risk of cancers is 
$\tilde{\epsilon}_i^{(j)}=10^{\epsilon_i^{(j)}}$, $j=1,2$. The mean value of $\tilde{\epsilon}_i^{(j)}$ is then $10^{ {\ln 10 \over 2} {\rm Var}[\epsilon_i^{(j)}]}, j=1,2$.
This shows that the magnitude of lifetime cancer risks not explained by the number of stem cell divisions is controlled only by the variance ${\rm Var}[\epsilon_i^{(j)}]$, for $j=1,2$.
Then, group 2 exhibits many more cancers than group 1 
($C_i^{(2)} \gg C_i^{(1)}$) in the following cases:
\begin{enumerate}
\item[(a)]  $\beta^{(2)} \gg \beta^{(1)}$ (much larger sensitivity to stem cell divisions) while ${\rm Var}[\epsilon_i^{(1)}]$ 
and ${\rm Var}[\epsilon_i^{(1)}]$ remain of the 
same order of magnitude;
\item[(b)] ${\rm Var}[\epsilon_i^{(2)}] \gg {\rm Var}[\epsilon_i^{(1)}]$, while the sensitivities $\beta^{(1)}$ and 
$\beta^{(2)}$ to stem cell divisions remain similar;
\item[(c)] $\beta^{(2)} \gg \beta^{(1)}$  and ${\rm Var}[\epsilon_i^{(2)}] \gg {\rm Var}[\epsilon_i^{(1)}]$.
\end{enumerate}
Consider the identity linking ${\rm Corr}[C_i^{(j)}, S_i^{(j)}]$
and ${\rm Var}[\epsilon_i^{(j)}]$ versus $\beta^{(j)}$ derived from (\ref{gbfwb}) and  (\ref{vfewfwc}),
\be
{\rm Corr}[C_i^{(j)}, S_i^{(j)}]  =  \left[ 1 + {{\rm Var}[\epsilon_i^{(j)}] \over (\beta^{(j)})^2 ~{\rm Var}[S_i]}\right]^{-{1 \over 2}}~.
\label{gbfrywb}
\ee
Case (a) leads to ${\rm Corr}[C_i^{(1)}, S_i^{(1)}] \ll {\rm Corr}[C_i^{(2)}, S_i^{(2)}]$, in contradiction with 
our assumption (\ref{rynehtbgw}).  Case (b) leads to ${\rm Corr}[C_i^{(1)}, S_i^{(1)}] \gg {\rm Corr}[C_i^{(2)}, S_i^{(2)}]$,
again in contradiction with (\ref{rynehtbgw}). In fact, expression (\ref{gbfrywb}) 
implies that ${\rm Corr}[C_i^{(j)}, S_i^{(j)}]$ remains unchanged when $\beta^{(j)}$ is increased arbitrarily while 
${\rm Var}[\epsilon_i^{(j)}]$ is also increased proportionally to $(\beta^{(j)})^2$, since ${\rm Var}[S_i]$ is assumed 
to be the same in the two groups. Thus, the assumption (\ref{rynehtbgw}) together with the identity
(\ref{gbfrywb}) imposes case (c) as the only general possibility
for $C_i^{(2)} \gg C_i^{(1)}$.


The analysis of Tomasetti and Vogelstein  \cite{Tomavogel15} does not distinguish
between groups exhibiting different cancer rates. This amounts to considering the total population of the two groups put together.
Then, in our hypothetical population, Tomasetti and Vogelstein would observe
\be
C_i^{(1)} + C_i^{(2)}= [\beta^{(1)} +  \beta^{(2)}]  S_i + \epsilon_i^{(1)} + \epsilon_i^{(2)}~,
\label{rtwhedyhbge2w}
\ee
using our assumption (\ref{fgngbw}).
In this meta-population, the correlation studied by Tomasetti and Vogelstein  \cite{Tomavogel15} 
is that between $C_i^{(1)} + C_i^{(2)}$ and $S_i$:
\be
{\rm Corr}[C_i^{(1)} + C_i^{(2)}, S_i] = { {\rm Cov}[C_i^{(1)}, S_i]  + {\rm Cov}[C_i^{(2)}, S_i] \over
\sqrt{ \left({\rm Var}[C_i^{(1)}] + {\rm Var}[C_i^{(2)}] + 2 \beta^{(1)}\beta^{(2)}  {\rm Var}[S_i]\right) {\rm Var}[S_i]}  }
\label{tjimi}
\ee
From (\ref{gbfwb}), (\ref{rgheythbwgr}), (\ref{rynehtbgw}) and (\ref{fgngbw}), we deduce
\be
{\rm Cov}[C_i^{(j)}, S_i] = \rho \sqrt{ {\rm Var}[C_i^{(j)}] {\rm Var}[S_i]}~,
\ee
which we insert in (\ref{tjimi}) to obtain
\be
{\rm Corr}[C_i^{(1)} + C_i^{(2)}, S_i] = \rho ~ { \sqrt{{\rm Var}[C_i^{(1)}]} + \sqrt{{\rm Var}[C_i^{(2)}] } \over
\sqrt{ {\rm Var}[C_i^{(1)}] + {\rm Var}[C_i^{(2)}] + 2 \beta^{(1)}\beta^{(2)}  {\rm Var}[S_i]}  }
\label{rthyy5hu6j}
\ee
By (\ref{vfewfwc}), we have
\be
{\rm Var}[C_i^{(j)}] \geq [\beta^{(j)}]^2  {\rm Var}[S_i]~~,
\label{guyjyhew}
\ee
which implies
\be 
{\rm Corr}[C_i^{(1)} + C_i^{(2)}, S_i]  \geq  {\rm Corr}[C_i^{(j)}, S_i] ~, ~~~~~j =1 ~{\rm or}~2~,
\label{rheruyjyhwg}
\ee
using definition (\ref{rynehtbgw}). 

The inequality (\ref{rheruyjyhwg}), which recovers a standard result in statistics, constitutes our main lever
to falsify Tomasetti and Vogelstein's claim: the correlation between stem cell divisions and cancer risks at the level of the total population 
is in fact no lower than that found at the individual group level.
In plain words, a strong correlation at the population level over all group types
is blind to the existence of strong differences in group susceptibilities to cancer
associated with other (i.e. environmental or hereditary) factors. In our hypothetical population,
one group shows a much higher cancer rate than the other, in the presence of a strong correlation between 
number of stem cell divisions and total cancer rate, but this does not allow one to conclude 
that the total number of stem cell divisions is the dominant
factor responsible for cancer in both groups (hence making cancer ``bad luck"). On the contrary, this result is compatible with a possibly strong influence from other environmental and genetic factors, here embodied in the variable $\epsilon_i^{(j)}$ as well as the
possible dependence of $\beta^{(j)}$ on the same factors.

We stress that our conclusion remains robust when relaxing the simple assumptions used in our hypothetical population.
For instance, the demonstration generalizes straightforwardly to more than two groups and even to a
continuum. The condition (\ref{rynehtbgw}) of equal correlations within the two groups can be generalized
to different values. And our argument and conclusion remain valid if
it would appear that the rate of divisions of the normal self-renewal stem cells may vary between groups.

A part of the conclusion that Couzin-Frankel \cite{Couzin-Frankel15} and Tomasetti and Vogelstein's \cite{Tomavogel15} draw
is thus unwarranted:  Tomasetti and Vogelstein's analysis does not allow 
one to conclude that the majority of cancers is due to unpreventable ``bad luck." 
We have just demonstrated that the existence of possibly strong differences in susceptibility to cancers, 
for instance due to environmental and genetic factors, has no effect on Tomasetti and Vogelstein's result 
that a large fraction of the variation in cancer risk among tissues, that is, differences in cancer incidence among different organs, can be explained by the number of stem cell divisions. Tomasetti and Vogelstein's 
findings point naturally to the prevalence of mutations during replications. This can explain why certain organs are more affected by cancer than others, but does not address the question of why certain populations or individuals are more
affected by cancer than others.

We have demonstrated that the coexistence of several populations with very different
cancer rates, for instance due to environmental and genetic causes, is compatible with the empirical evidence
of a strong correlation between the total number of cell divisions and cancer risks \cite{Tomavogel15}.
One may ask whether our hypothetical population made of two groups with
$\beta^{(2)} \gg \beta^{(1)}$  and ${\rm Var}[\epsilon_i^{(2)}] \gg {\rm Var}[\epsilon_i^{(1)}]$ (case (c))
has anything to do with reality?
The answer is empirical and requires to extend Tomasetti and Vogelstein's analysis to 
different cohorts under various environmental stressors as in the Framingham Heart Study of NIH \cite{Framingham},
the China-Cornell-Oxford Project \cite{CampbellCampbell06} and others  \cite{Cairns86, Pisani02, Calle03, Montesano01}.
Case (c) corresponds to a consistently large correlation between number of stem cell divisions and cancer risk and
provides an interesting testable hypothesis, namely that controllable environmental factors and/or genetic traits
impact both the cancer risks related to stem cell divisions and those that seem unrelated
to stem cell divisions. This requires to study {\it conditional} correlations, thus extending
the unconditional correlation study of Tomasetti and Vogelstein (since no condition on separate groups
or cohorts is imposed in their study).

Indications of strong environmental factors are actually observed in figure 1 of Ref.\cite{Tomavogel15}:
(i) lifetime lung cancer risk is multiplied by 12 by smoking; (ii) lifetime head and neck cancer risk is multiplied by 6 
after Human papillomavirus contamination; (iii) Hepatocellular carcinoma risk is multiplied by 10
after hepatitis C virus contamination; (iv) colorectal cancer risk is multiplied by 12 in the presence of
familial adenomatous polyposis.  A possible source of confusion may be due to
the existence of more than 200 different kinds of cancers according to present taxonomy, with 
many more subtypes coming in month by month.
For the well-known cancer types, epidemiology shows a strong link between environmental and life style factors.
For the many other so-called sporadic cancers, epidemiological studies are much less advanced.
We hope that the present note will help refocus on the importance of 
environmental  and predisposing genetic factors  \cite{Lichtenstein00,Lanzmann09,CampbellCampbell06,Servan-Schreiber}
and not miss the forest for the trees.

We acknowledge very helpful feedbacks from Thomas Cerny, Jean-Yves Henry, and Christine Sadeghi.

\end{document}